\documentclass[aps,pre,twocolumn,showpacs,amsfonts,amssymb,floatfix]{revtex4}
\usepackage{graphics}
\usepackage{psfrag}
\usepackage{placeins}
\usepackage{graphicx}
\usepackage{amssymb}
\usepackage{amsmath}
\newcount\hour \newcount\minute
\hour=\time \divide \hour by 60 \minute=\time \count99=\hour
\multiply \count99 by -60 \advance\minute by \count99 \hour=\time
\divide \hour by 60
\begin{document}

\title{Generation of spatiotemporal correlated noise in 1+1 dimensions}
\author{Arne Traulsen, Karen Lippert, and Ulrich Behn}
\affiliation{Institut f\"ur  Theoretische Physik, Universit\"at Leipzig, Vor dem
Hospitaltore 1, 04103 Leipzig, Germany}
\begin{abstract}

We propose a generalization of the Ornstein-Uhlenbeck process in $1+1$
dimensions which is the product of a temporal Ornstein-Uhlenbeck process with 
a spatial one and has exponentially decaying autocorrelation.  The generalized 
Langevin equation of the process, the corresponding Fokker-Planck equation, and
a discrete integral algorithm for numerical simulation is given. The process is
an alternative to a recently  proposed spatiotemporal correlated model process
[J. Garc\'\i a-Ojalvo et al., Phys.\ Rev.\ A {\bf 46}, 4670 (1992)] for which we
calculate explicitely the hitherto not known autocorrelation function in real
space.

\end{abstract}
\maketitle

\section{Introduction}

Noise induced phenomena are subject of considerable recent attention
\cite{HL84,GS99}. After considering in the early phase systems with only few
degrees of freedom \cite{HL84} in the last decade the effects of noise in
spatially distributed systems have been investigated \cite{GS99}. In this
context stochastic model processes are necessary to mimick spatiotemporal
fluctuations of different origin. If characteristic time and length scales of
system and noise are clearly separated, the use of a spatiotemporal Gaussian
white noise may be justified, but it can also lead to spurious results,
\cite{Honda96,Honda97,LS98}. There are physical situations where the
characteristic scales are not well separated, e.g. in externally driven systems
\cite{GS94}, or where the square of the driving stochastic process is involved.
Both is the case in  electrohydrodynamic convection in nematic liquid crystals,
cf.\ e.g. \cite{PB98}, driven by external stochastic electric fields 
\cite{BLJ98,JSB99}. Further examples are the influence of spatiotemporal
colored noise on spatiotemporal chaos modeled by the complex Ginzburg-Landau
equation \cite{WO02} and on networks of excitable systems displaying
spatiotemporal stochastic resonance \cite{BK03}. This sufficiently motivates to
model correlated spatiotemporal fluctuations.  For an approach based on
different grounds see, e.g., \cite{VMR94}. 

A frequently used spatiotemporal correlated model process was introduced by
Garc\'\i a-Ojalvo et al.\ (GSR) \cite{GSR92} who considered in spatial
dimension $d$, ${\bf r} \in R^d$ the stochastic partial differential equation
(PDE)
\begin{equation} \label{GO}
\tau \frac{\partial }{ \partial t} \varphi({\bf r},t) = - \left( 1 - \lambda ^2
\Delta \right)\varphi({\bf r},t) +\xi({\bf r},t), 
\end{equation} 
where the additive driving process $\xi({\bf r},t)$ is Gaussian distributed
with zero mean and with autocorrelation $ K^{\xi}({\bf r},t;{\bf
r}',t')=\langle \xi({\bf r},t) \xi({\bf r}',t')\rangle =\sigma^2\delta({\bf
r}-{\bf r}') \delta(t-t')$. 

The heuristics of Eq. (\ref{GO}) is evident: the diffusive term  effectively
reduces the life-time of Fourier components with wavelengths short compared to
$\lambda$, see also below. For $\lambda =0$ it reduces to the Langevin equation
defining the common temporal Ornstein-Uhlenbeck process (OUP),
\cite{UO30,Chandrasekhar43,WU45}.  The solutions are thus, in a sense,
generalizations of the OUP.   Equation (\ref{GO}) belongs to a class of
stochastic PDEs for which existence and uniqueness of the solution are proven
rigorously \cite{GP93}. It is discussed  also in the context of
reaction-diffusion systems, see e.g. \cite{Gardiner}, and within a generating
functional approach \cite{VMR94}.

In this paper we propose an alternative spatiotemporal generalization of the
OUP in 1+1 dimension which is simply the product of a temporal OUP with a
spatial one and has exponentially decaying autocorrelation. To make the paper
self-contained and to introduce the notation which is  used in the sequel we
shortly recall in Sec.\ II basic properties and numerical generation of the
common   OUP in one (temporal) dimension. The  scaling necessary when
transforming between discrete and continuous formulations is carefully
discussed. In Sec.\ III the generalized OUP is constructed independently within
a spatially discretized scheme and in a continuous version as the solution of a
stochastic PDE different from Eq.\ (\ref{GO}). Subsequently, conditions which
ensure stationarity and homogeneity are discussed, the generalized
Fokker-Planck equation and its stationary solution are given, and numerically
generated data are compared with the analytic results. In Sec.\ IV the
autocorrelation function of the GSR process is explicitely calculated in real
space for $d=1$ in both  continuous and discrete formulation and compared with
numerical results. Previous work studied the behaviour in real space only for
spatial dimensions $d=2$, \cite{GSR92,LB93},  and $d=3$, \cite{LB93},  cf.\
however \cite{LB93mistakes}. Contrary to the folklore \cite{SGG98,WO02},
the autocorrelations of the GSR process decay  not exponentially but in a more
intricate way.\\ Problems connected with the generalization to higher
dimensions are shortly discussed in the concluding section.

\section{The Ornstein-Uhlenbeck Process}
\label{Ornstein-Uhlenbeck-Process}

The OUP ist the only stationary Gaussian Markov process
with exponentially decaying autocorrelation (Doob's Theorem \cite{Doob42}).
Realizations $\eta(t)$ of the OUP can be generated solving the Langevin
equation
\begin{equation}
\label{langevin}
\tau \frac{d}{dt}\eta(t) = -\eta(t) +  \xi (t),
\end{equation}
where $\xi(t)$ is a Gaussian white noise with $\langle \xi(t) \xi(t') \rangle =
\sigma_{\rm{t}}^2 \delta(t-t')$. In mathematically precise form Eq.
(\ref{langevin}) reads
\begin{equation}
\label{sde}
\tau d \eta(t) = -  \eta(t) dt + dW(t),
\end{equation}
where $W$ is a Wiener process with $\langle W(t) W(t') \rangle=
\sigma_{\rm{t}}^2 \min(t,t')$; note $dW(t)/dt=\xi(t)$. Solving Eq.
(\ref{sde}) with initial condition $ \eta(t_0)= \eta_0 $ gives
\begin{equation} \label{OUPgeneralsolution}
\eta(t)=\eta_0 e^{-(t-t_0)/\tau} +\frac{1}{\tau} e^{-t/\tau}
\int_{t_0}^t dW(s)e^{s/\tau},
\end{equation}
which has the autocorrelation
\begin{equation}
\langle \eta(t) \eta(t') \rangle = \left(\langle \eta_0^2 \rangle -
\frac{\sigma_{\rm{t}}^2}{2 \tau}\right) e^{-(t+t'-2t_0)/\tau}
+ \frac{\sigma_{\rm{t}}^2}{2 \tau} e^{-|t-t'|/\tau}.
\end{equation}
The process becomes stationary if the initial values are Gaussian distributed
with zero mean and variance $ \sigma_{\rm{t}}^2/2 \tau$, or in the limit $t,t'
\rightarrow \infty$, or for $t_0 \rightarrow -\infty$; it is then the OUP. We
denote the stationary part of the autocorrelation function as
\begin{equation}
\label{auto}
K^{\eta}(t-t') = \frac{\sigma_{\rm{t}}^2 }{ 2 \tau} e^{-|t-t'| /\tau }\; .
\end{equation}
Naturally, $K^{\eta}(t-t')$ solves the inhomogeneous equation which is
obtained by multiplying Eq. (\ref{langevin}) with $\eta(t)$ given by
(\ref{OUPgeneralsolution}) and averaging
\begin{equation}\label{inhDGLKOUP}
\tau \frac{d}{dt}K^{\eta}(t-t')=-K^{\eta}(t-t')+\Theta(t'-t)\frac
{\sigma^2}{\tau}e^{-(t'-t)/\tau}.
\end{equation}

Given the value $\eta(t)$ we can obtain $\eta(t+\Delta t)$ as
\begin{equation}
\label{OUPincrementsolution}
\eta(t+\Delta t) = \eta(t) e^{-\Delta t/\tau} +\frac{1}{\tau}
e^{-(t+\Delta t)/\tau} \int_t^{t+\Delta t} dW(s) e^{s/\tau},
\end{equation}
where the last term of the r.h.s is a stochastic increment. The increments in
non-overlapping time intervals are obviously independent; they have zero mean
and the variance, cf.\ e.g.\ \cite{FGRV88},
\begin{equation}
\frac{\sigma_{\rm{t}}^2}{2 \tau} \left(1-e^{-2 \Delta t/\tau} \right).
\end{equation}
Introducing the notation $\eta_t=\eta(\Delta t \cdot t)$, where $\Delta t$ is
fixed and $t=\ldots,-1,0,1,\ldots$, one obtains, for every choice of $\Delta
t$, an exact recursion relation for equidistant discrete times (discrete
integral algorithm),
\begin{equation}
\label{OUPFoxsolution}
\eta_{t+1}=\eta_t \;e^{- \Delta t /\tau} +\frac{\sigma_{\rm{t}}}{\sqrt{2 \tau}}
\sqrt{1-e^{-2 \Delta t /\tau}}\; \tilde{\xi}_{t+1},
\end{equation}
where $\tilde{\xi}_{t}$ are independent zero mean Gaussian random numbers
with variance one, cf.\ \cite{Gillespie96}.

For {\it small} $\Delta t$ a Taylor expansion of the r.h.s.\ of  Eq.\
(\ref{OUPincrementsolution})  leads to
\begin{equation}
\label{OUPTaylorsolution}
\eta(t+\Delta t)=\eta(t) \;\left(1-\frac{\Delta t}{\tau} \right) + \frac{\Delta
t}{\tau}\xi(t+\Delta t).
\end{equation}

The discrete version of Eq. (\ref{OUPTaylorsolution}) is obtained using the
above notation and replacing $\xi(t)$ by $\sigma_{\rm{t}}
\tilde{\xi}_t/\sqrt{\Delta t}$. This ensures the correct autocorrelation in the
continuum limit observing $\lim_{\Delta t \to 0}  \delta_{t,t'}/{\Delta
t}=\delta(t-t')$ and amounts to a rescaled  variance 
$\tilde{\sigma}_{\rm{t}}^2=\sigma_{\rm{t}}^2/\Delta t$,
\begin{equation}
\label{OUPdiscretesolution}
\eta_{t+1}=\eta_t \left(1-\frac{\Delta t}{\tau} \right) + \frac{\Delta
t}{\tau} \tilde\sigma_{\rm{t}}\tilde{\xi}_{t+1}.
\end{equation}
All results are consistent: Eq.\ (\ref{OUPTaylorsolution}) can be derived from
Eq.\ (\ref{langevin}) using an Euler discretization, and Eq.\
(\ref{OUPdiscretesolution}) from Eq.\ (\ref{OUPFoxsolution}) by a Taylor
expansion of the coefficients.

\section{A Generalization to 1+1 Dimensions}
\label{Generalization}

In this Section we generalize the OUP and construct in $1+1$ dimensions a
spatiotemporal random field $\varphi (x,t)$ with reasonable properties. For
fixed $x$ the process should be the common temporal OUP described above
and for fixed $t$ a spatial OUP. It is reasonable to require translational invariance,
analogous to temporal stationarity, of all averages  and an exponential decay
of the spatiotemporal autocorrelation
\begin{eqnarray}
\label{xtkorr}
& & K^{\varphi}(x-x',t-t') =  \langle \varphi(x,t) \varphi(x',t') \rangle  \nonumber \\
& & =\frac{\sigma^2}{ 4 \lambda \tau} e^{ -|x-x'| / \lambda -|t-t'| / \tau},
\end{eqnarray}
where  $\sigma=\sigma_{\rm{s}} \sigma_{\rm{t}}$,  $\sigma_{\rm{s}}$
and $\lambda$ characterizing for fixed time the spatial process.

We propose two independent schemes leading to the same result. First we employ
a spatially  discrete scheme to construct more general spatiotemporal
correlated processes given in \cite{WCW95}. Alternatively, we consider a linear
stochastic PDE different from Eq. (\ref{GO}) driven by additive Gaussian
spatiotemporal white noise and show that it's stationary solutions are Gaussian
distributed and have the desired properties. The analytic results are
corroborated by numerical data.

\subsection{Recursive Generation}

We consider the field $\varphi(x,t)$ on equidistant lattice sites $i$,  $\; i =
0,\ldots, N$, adopting the notation $\varphi_i(t) = \varphi(\Delta x \cdot
i,t)$. In the first step of construction we generate spatially independent
OUPs $\eta_i(t)$ using a standard algorithm, e.g.
\cite{FGRV88,MP89}, with autocorrelation
\begin{equation}
K_{ij}^{\eta}(t-t')=\langle \eta_i(t) \eta_{j}(t') \rangle =
\frac{{\sigma}^2}{4 \lambda \tau} e^{-|t-t'|/\tau} \delta_{i,j}.
\end{equation}
We then, as proposed in \cite{WCW95} to construct a more general spatiotemporal
correlated noise,  superpose these processes
\begin{equation}
\varphi_{i}(t) = \sum_{k=0}^i a_{ik} \eta_{k}(t),\; i=0,\ldots, N.
\label{ansatz}
\end{equation}
Since this expression is linear in $\eta_k$ also  the $\varphi_i$ are Gaussian
distributed with zero mean. Requiring that the spatial autocorrelation is the
discrete version of Eq. (\ref{xtkorr}) for equal times,
\begin{equation}
K^{\varphi}_{ij}(0)
= \sum_{k=0}^{\min(i,j)} a_{ik}a_{jk}\langle \eta_k(t)^2 \rangle
= \frac{{\sigma}^2}{ 4 \tau \lambda} e^{-|i-j|\Delta x / \lambda},
\end{equation}
determines the coefficients $a_{jk}$. It is easy to check that
\begin{equation}
a_{jk} =e^{-(j-k) \Delta x /\lambda}
\left(\sqrt{1-e^{-2 \Delta x/\lambda}}\right)^{1-\delta_{k,0}},
\end{equation}
where $j=0,\ldots, N$ and $k=0,\ldots, j$. Using these coefficients we write
\begin{eqnarray}
\label{scheme1}
& & \varphi_j (t) = e^{-j\Delta x / \lambda} \eta_0(t) \nonumber\\
& & + e^{-(j-1)\Delta x / \lambda} \sqrt{1 - e^{-2\Delta x / \lambda}}\eta_1(t)
 \nonumber \\
& & + \ldots +\sqrt{1 - e^{-2\Delta x / \lambda}}\eta_j(t).
\end{eqnarray}
With the corresponding formula for $ \varphi_{j+1} (t)$ we find 
\begin{equation}\label{spatint}
\varphi_{j+1}(t)= e^{-\Delta x/\lambda}\varphi_j(t)
+\sqrt {1- e^{-2\Delta x/\lambda}}\eta_{j+1}(t),
\end{equation}
where $\eta_{j+1}(t)$ are the spatially independent random numbers specified
above, each of which being a temporal OUP. Obviously, this is the spatial
analogue of the discrete integral algorithm (\ref{OUPFoxsolution}) for the
temporal process.

In discrete notation both for space and time we insert $\eta_{j+1,t+1}$ from Eq.
(\ref{OUPFoxsolution}), after replacing $ {\sigma_{\rm{t}}}/{\sqrt{2 \tau}}
\rightarrow {\sigma}/{\sqrt{4 \tau \lambda}}$, into Eq.
(\ref{spatint}) (written for $\varphi_{j+1,t+1}$) and obtain finally 
\begin{eqnarray}
\label{intalg}
& & \varphi_{j+1,t+1} = e^{-\Delta x / \lambda} \varphi_{j,t+1}
+ e^{-\Delta t / \tau } \varphi_{j+1,t} \nonumber    \\
& & - e^{-\Delta t / \tau - \Delta x / \lambda } \varphi_{j,t}  \nonumber \\
& & + \sqrt{1-e^{-2\Delta x / \lambda}} \sqrt{1-e^{-2\Delta t / \tau}}
 \frac{{\sigma} }{ \sqrt { 4 \tau \lambda}} \tilde{\xi}_{j+1,t+1},
\end{eqnarray}
where $ \langle \tilde{\xi}_{i,t} \tilde{\xi}_{j,t'} \rangle =\delta_{i,j}
\delta_{t,t'} $. For simulations this  discrete integral algorithm is 
preferrable since by construction it is correct for any choice of $\Delta x$ and
$\Delta t$.

Expanding the coefficients in Eq. (\ref{intalg}) for small $\Delta x$ and
$\Delta t$ a first order discrete differential algorithm, the generalization
of Eq. (\ref{OUPdiscretesolution}), is obtained,
\begin{eqnarray}
\label{difalgdisc}
& & \varphi_{j+1,t+1} =  \left (1-\frac{\Delta x }{ \lambda} \right ) \varphi_{j,t+1}
 + \left (1-\frac{\Delta t }{ \tau }\right ) \varphi_{j+1,t} \nonumber \\
& & -\left (1-\frac{\Delta x }{ \lambda} \right )
\left (1-\frac{\Delta t }{ \tau} \right )\varphi_{j,t} \nonumber \\
& & +\frac{\Delta x \Delta t}{ \lambda \tau} \tilde{\sigma} \tilde{\xi}_{j+1,t+1},
\end{eqnarray}
where ${\tilde{\sigma}}=\sigma/\sqrt{\Delta x \Delta t}$.
Writing Eq. (\ref{difalgdisc}) in continuous notation,
\begin{eqnarray}
\label{difalgcont}
& & \varphi(x+\Delta x, t+\Delta t) =  \left (1-\frac{\Delta x }{ \lambda} \right )
\varphi (x,t+\Delta t) \nonumber \\
& & + \left (1-\frac{\Delta t }{ \tau }\right )
\varphi (x+\Delta x,t) \nonumber \\
& & -\left (1-\frac{\Delta x }{ \lambda} \right ) \left (1-\frac{\Delta t }{ \tau}
\right )\varphi (x,t) \nonumber \\
& & +\frac{\Delta x \Delta t }{ \lambda \tau} \xi(x+\Delta x, t+\Delta t),
\end{eqnarray}
we have replaced $\tilde{\sigma} \tilde{\xi}_{j,t} \to \xi(x,t)$ in
complete analogy with the rescaling for the temporal OUP. The autocorrelation
of the spatiotemporal Gaussian white noise is $\langle \xi(x,t)
\xi(x',t')\rangle=\sigma^2 \delta(x-x')\delta(t-t')$.

\subsection{Continuous Approach}

\subsubsection{Generalized Langevin equation}

An alternative approach starts from a generalized Langevin equation in 1+1
dimensions, the stochastic PDE
\begin{equation}
\label{basiceq}
 \left ( 1 + \tau \frac{\partial }{ \partial t} + \lambda
\frac{\partial }{ \partial x} + \lambda \tau \frac{\partial ^2}{ \partial x
\partial t}\right ) \varphi(x,t) = \xi(x,t). 
\end{equation}
which for $\lambda=0$ reproduces the Langevin equation (\ref{langevin})
generating the temporal OUP and for $\tau=0$ that of the spatial OUP. The
spatiotemporal Gaussian white noise can be conceived as the product
$\xi(x,t)=\xi(x)\xi(t)$, where $\xi(x)$ and $\xi(t)$ denote independent spatial
and temporal Gaussian white noise, respectively. 

It is interesting to note that  Eq.\ (\ref{basiceq}) is a hyperbolic PDE whereas
Eq.\ (\ref{GO})  is a parabolic one. Eq.\ (\ref{basiceq}) has the two families of
characteristics $x=$ const and $t=$ const, the latter one is the only family of
characteristics of Eq.\ (\ref{GO}). Correspondingly, the solution of Eq.\
(\ref{GO}) reproduces in the limit $\lambda \to 0$ the temporal OUP multiplied
by $\delta(x-x')$ but $\tau \to 0$ results not in the spatial OUP, see
below.

The stochastic PDE (\ref{basiceq}) can be obtained from the continuous
differential algorithm  (\ref{difalgcont}) by Taylor expansion of $\varphi$ for
small $\Delta x$ and $\Delta t$ and performing the limit $\Delta x, \Delta t
\to 0$. Alternatively, it can be conceived as the product of
the two Langevin equations for a temporal OUP, Eq.\ (\ref{langevin}), and its
spatial analogue. For this we denote the product of the temporal and the
spatial OUPs by $\varphi(x,t)$ and observe that the differential operator on
the l.h.s.\ of  Eq.\  (\ref{basiceq}) factorizes as $\left ( 1 + \tau {\partial
}/{ \partial t} + \lambda {\partial }/{ \partial x} + \lambda \tau {\partial
^2}/{ \partial x \partial t}\right )=\left(1+ \tau{\partial}/{\partial t}
\right) \left(1+ \lambda{\partial}/{\partial x} \right)$.   

Using a separation ansatz, a solution of Eq.\ (\ref{basiceq}) can be written as
\begin{gather}\label{phi}
\varphi(x,t)=f(x)g(t), \;\;\;\text{where}\\ \label{f}
f(x)=f_0 e^{-(x-x_0)/\lambda} +\frac{A}{\lambda} e^{-x/\lambda}
\int_{x_0}^x dW(y)e^{y/\lambda},\\
\label{g}
g(t)=g_0 e^{-(t-t_0)/\tau} +\frac{1}{A\tau} e^{-t/\tau}
\int_{t_0}^t dW(s)e^{s/\tau},
\end{gather}
and $f(x_0)=f_0$ and $g(t_0)=g_0$ denote boundary and initial values. The
initial and boundary processes $\varphi (x,t_0)$ and $\varphi (x_0,t)$ are OUPs
with correlation length $\lambda$ and correlation time $\tau$, respectively.
Note the appearance of the extra factors $A$ and $1/A$ in Eqs.\ (\ref{f}) and
(\ref{g}) respectively,  compared with the process given by Eq.\
(\ref{OUPgeneralsolution}).  $0<|A|<\infty $ is an arbitrary constant which
corresponds to the separation constant for a deterministic PDE. In the
nonstationary case it weights the relative influence of the initial and
boundary realizations. In the term of $\varphi(x,t)$ which survives in the
stationary case, $A$ cancels and naturally its value plays no role, see below. 

Exploiting that the spatial and the temporal Wiener processes $W(y)$ and $W(s)$
are independent and have zero mean we obtain the autocorrelation function
\begin{eqnarray}
\label{Correlation}
& & K^{\varphi}(x,t;x',t') \nonumber \\
& & = \frac{\sigma^2 }{ 4  \tau \lambda} e^{-|x-x'| /
\lambda-|t-t'|/ \tau} \nonumber \\
& &+
\frac{\sigma_{\rm{t}}^2 }{ 2 \tau}
\left( \langle f_0^2 \rangle \frac{1}{ A^2}- \frac{\sigma_{\rm{s}}^2 }{ 2 \lambda} \right)
e^{-(x + x'-2 x_0)/\lambda-|t-t'| / \tau} \nonumber  \\
& &+
\frac{\sigma_{\rm{s}}^2}{ 2 \lambda}
\left( \langle g_0^2 \rangle  A^2 - \frac{ \sigma_{\rm{t}}^2 }{ 2 \tau} \right)
e^{-|x-x'|/ \lambda-(t +t' - 2 t_0 )/ \tau} \nonumber \\
& &+
\left( \langle f_0^2 \rangle \frac{1}{ A^2}- \frac{\sigma_{\rm{s}}^2 }{ 2 \lambda} \right)
\left( \langle g_0^2\rangle  A^2 - \frac{ \sigma_{\rm{t}}^2 }{ 2 \tau} \right) \nonumber \\
& & \times
e^{-(x +x'-2 x_0)/ \lambda -(t + t' -2 t_0)/ \tau}.
\end{eqnarray}

The first term on the r.h.s.  is  just the desired stationary and homogeneous
autocorrelation, independent on the boundary and initial conditions, cf.\ Eq.\
(\ref{xtkorr}). The remaining terms disappear for $x_0 \to -\infty$ and $t_0
\to -\infty$, respectively. A second possibility to make the nonstationary and
nonhomogeneous terms vanish is to chose $f_0$ and $g_0$ as zero mean Gaussian
distributed with variance  such that 
\begin{equation}
\label{LBbrackets}
\langle f_0^2 \rangle =  \frac{\sigma_{\rm{s}}^2}{2 \lambda} A^2,
\qquad \hbox{and} \qquad
\langle g_0^2 \rangle =  \frac{\sigma_{\rm{t}}^2}{2 \tau} \frac{1}{A^2}.
\end{equation}
In this case the process will be homogeneous and stationary from the
beginning.   The variance of the process $\varphi$ is independent of
$(x,t)$, $\langle \varphi^2(x,t) \rangle = \langle f_0^2\rangle \langle
g_0^2\rangle = \sigma^2/4 \lambda \tau$.

\subsubsection{Generalized Fokker-Planck Equation}

The Fokker-Planck equation (FPE) corresponding to a stochastic PDE should be a
functional equation. For the spatially discretized system the FPE is a matrix
equation. We will derive for this case the generalized FPE and its stationary
solution.  Discretizing Eq.\ (\ref{basiceq}) using a first order Euler-scheme
gives the system of ordinary differential equations,
\begin{equation}
\label{LBmateq}
\tau {\bf C} d {\boldsymbol \varphi}(t) = -{\bf C} {\boldsymbol \varphi}(t) dt +
\frac{\sigma_{\rm{s}}}{\sqrt{\Delta x}} d {\bf W}(t),
\end{equation}
where ${\boldsymbol \varphi}=(\varphi_1,\ldots,\varphi_N)^{\rm T}$
and  ${\bf W}=(W_1,\ldots,W_N)^{\rm T}$.
The matrix $\bf C$ has the non-vanishing elements
\begin{equation}
\label{LBMatEl}
c_{i,i} = c_0 = 1+\frac{\lambda }{ \Delta x}
\hbox{,} \quad
c_{i+1,i}= c_1 = -\frac{\lambda }{ \Delta x},
\end{equation}
Since $\det {\bf C} \not= 0$ we can multiply Eq.\ (\ref{LBmateq}) with
$ {\bf C}^{-1}$ and obtain
\begin{equation}
\label{LBmultOUP}
\tau d {\boldsymbol \varphi}(t) = -{\boldsymbol \varphi}(t) dt +
\frac{\sigma_{\rm{s}}}{\sqrt{\Delta x}} {\bf C}^{-1} \,d {\bf W}(t).
\end{equation}
Now we can treat the system as a multivariate OUP. 
It can be shown \cite{Gardiner} that the corresponding FPE is
\begin{eqnarray}
\label{LBFokkerPlanckMOUP}
& &\frac{\partial}{\partial t}p = -\sum_{i} \frac{\partial}{\partial \varphi_i}
\sum_j\Big[ -\frac{1}{\tau}\delta_{ij} \varphi_j p  \nonumber \\
& &-\frac{\sigma_{\rm{s}}^2 \sigma_{\rm{t}}^2}{2 \tau^2 \Delta x}
\left({\bf C}^{-1} \left({\bf C}^{-1}\right)^{\rm T}\right)_{ij}
\frac{\partial}{\partial \varphi_j}p \Big]    \nonumber \\
& & = -\sum_{i} \frac{\partial}{\partial \varphi_i} J_i
= - \nabla_{\varphi} {\bf J},
\end{eqnarray}
where $p=p({\boldsymbol \varphi},t|{\boldsymbol \varphi_0},t_0)$ is the
transition probability density and $\bf J$ is the hereby defined probability
current density. We note that in our case ${\bf C}^{-1}\left({\bf
C}^{-1}\right)^{\rm T} = \left({\bf C}^{\rm T} {\bf C}\right)^{-1}$.

A stationary solution of Eq.\ (\ref{LBFokkerPlanckMOUP}), means ${\bf J} = {\rm
const}$. For natural boundaries where the probability current vanishes we have
\begin{equation}
\label{LBprobabilitycurrent}
J_i =  \sum_j\left[
-\frac{1}{\tau}\delta_{ij} \varphi_j p_s - \frac{\sigma^2}{2 \tau^2 \Delta x}
\left({\bf C}^{\rm T} {\bf C}\right)^{-1}_{ij}
\frac{\partial}{\partial \varphi_j}p_s \right] = 0.
\end{equation}
From Eq.\ (\ref{LBprobabilitycurrent}) we get
\begin{equation}
\label{LBFPGradient}
\frac{\partial}{\partial \varphi_l} \ln p_s  =
\sum_{j}\left[ -\frac{2 \tau \Delta x}{ \sigma^2}\left({\bf C}^{\rm T}
{\bf C}\right)_{lj}  \varphi_j  \right].
\end{equation}
Now the non-vanishing elements of ${\bf C}^{\rm T}{\bf C}$ can be computed
from Eq.\ (\ref{LBMatEl}) as
\begin{equation}
\left({\bf C}^{\rm T}{\bf C}\right)_{i,i}=c_{0}^2+c_{1}^2
\hbox{,} \quad
\left({\bf C}^{\rm T}{\bf C}\right)_{i,i\pm 1}= c_{0}c_{1}.
\end{equation}
As the right hand side of Eq.\ (\ref{LBFPGradient}) is a gradient
(${\bf C}^{\rm T}{\bf C}$ is symmetric), the potential conditions are fulfilled
and a simple integration gives
\begin{eqnarray}\label{gauss}
p_s({\boldsymbol \varphi}) = {\cal N}\exp\left[-\frac{\tau \Delta x}{ \sigma^2}
{\boldsymbol \varphi}^{\rm T}{\bf C}^{\rm T}{\bf C} {\boldsymbol \varphi}  \right],
\end{eqnarray}
where ${\cal N}$ is the normalization factor. ${\bf C}^{\rm T}{\bf C}$ is
an oscillation matrix \cite{Gantmacher} with the positive eigenvalues
\begin{eqnarray}
\Lambda_j =
1+ 2 \left( \frac{\lambda}{\Delta x} +
\frac{\lambda^2}{\Delta x^2}\right)
\left(1-\cos\left(\frac{\pi j}{N+1}\right)\right).
\end{eqnarray}
Thus the stationary solution can be normalized,  ${\cal N} = \prod_{j=1}^N
\sigma \left( \pi \tau \Delta x \Lambda_j\right)^{-1/2}$, and  the stationary
probability density is indeed the  zero mean Gaussian distribution
(\ref{gauss}).

\subsection{Comparison with Numerics}

\begin{figure}[t]
\centering
\scriptsize
\psfrag{a}{\bf (a)}
\psfrag{b}{\bf (b)}
\psfrag{0.1}{0.1} \psfrag{ma}{1.0}
\psfrag{y}[][][1][180]{$K^{\varphi}(x-x',0)$}
\psfrag{x}{$x-x'$} \psfrag{0}{0} \psfrag{50}{} \psfrag{100}{100}
\psfrag{150}{} \psfrag{200}{200} \psfrag{250}{} \psfrag{300}{300}
\psfrag{0.001}{0.001} \psfrag{0.01}{0.01} \psfrag{0.1}{0.1}
\includegraphics[totalheight=4.5cm]{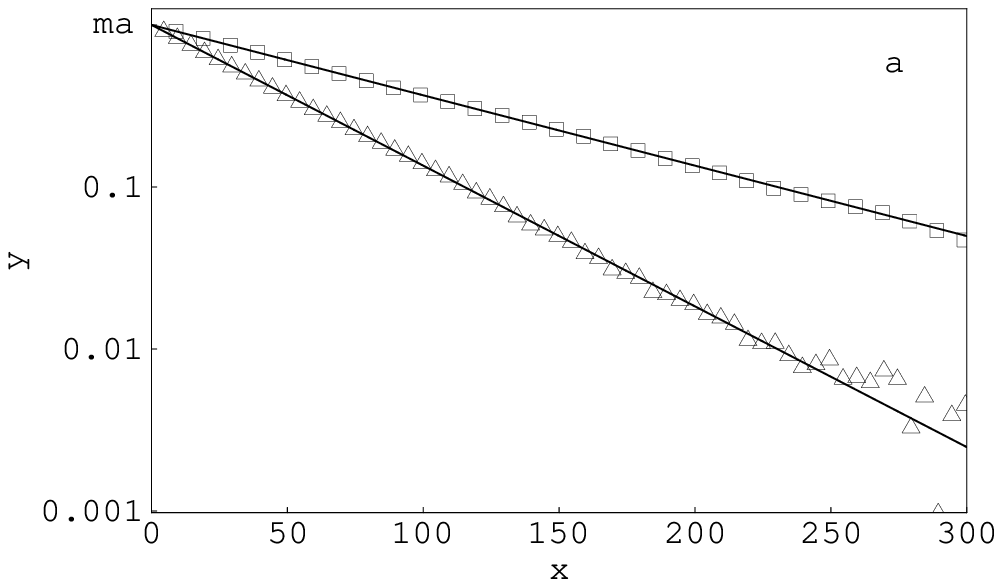}
\hfill
\psfrag{y}[][][1][180]{$K^{\varphi}(0,t-t')$}
\psfrag{t}{$t-t'$}
\includegraphics[totalheight=4.5cm]{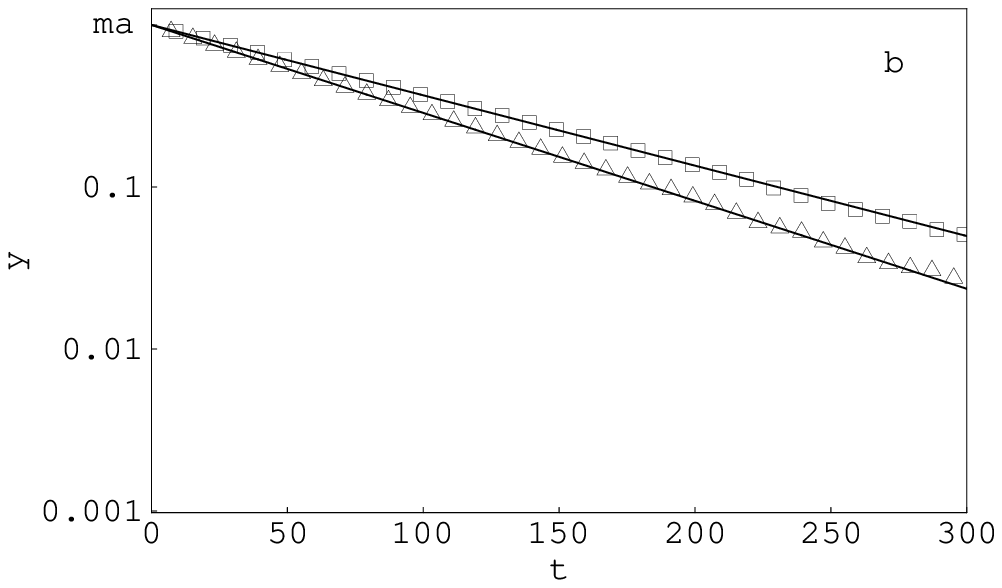}
\caption{Autocorrelation of the generalized OUP in 1+1 dimensions normalized by
the variance $\tilde{\sigma}^2/4 \lambda \tau$. Comparison of  analytical and
numerical results for (a) fixed temporal $(t=t'=100)$ and (b) fixed spatial
$(x=x'=100)$ argument. The lines show the analytic results from Eq.\
(\ref{Correlation}), the symbols are  the results of simulations (squares:
$\lambda=\tau=100$, triangles: $\lambda=50$, $\tau=80$). Stationarity was
ensured imposing the corresponding initial and boundary processes, see text.
Averages over $10^5$  realizations (N=1000, $\tilde{\sigma}=1$, $ \Delta
x=\lambda / 100$, $\Delta t=\tau / 100$).}
\label{LBAlgGrafik}
\normalsize
\end{figure}

We compare the analytically given autocorrelation with numerically generated
data obtained with the discrete integral algorithm provided by Eq.\
(\ref{intalg}). Fig. \ref{LBAlgGrafik} shows a good agreement for fixed
temporal and fixed spatial argument, respectively, imposing initial and
boundary conditions which ensure stationarity and homogeneity as described
above.

We also determined the mean square deviation of the variance of averages over
$10^5$ independent realizations which is governed by the $\chi^2$-distribution.
The variance was always found within a 80 \% confidence interval.

\section{The Approach of GSR in 1+1 Dimensions}
\label{GOApproach}

The above proposed generalization of the OUP has {\it by construction}
autocorrelations decaying exponentially both in space and time. This is in
contrast to the spatiotemporal correlated noise proposed by GSR \cite{GSR92}.
Since the autocorrelation for 1+1 dimensions in real space was not explicitly
calculated in the previous literature we below consider this case. Again, we
derive the result in a continuous approach and in a spatially discretized
scheme and compare the analytical  results with numerical data. The
autocorrelation in real space for spatial dimension $d  \geq 2$ is evaluated in
a different context in \cite{TB}. In reciprocal space,  the result is given for
general $d$ in \cite{LB93,VMR94}, cf.\ also \cite{LB93mistakes}.

\subsection{Continuous Approach}

We start with the Fourier transform of Eq.\ (\ref{GO}) in $d=1$ which reads 
\begin{equation}
\label{GOFourier}
\tau \frac{\partial }{ \partial t}\varphi(k,t) = -c(k) \varphi(k,t) + \xi(k,t),
\end{equation}
where $c(k)=1+\lambda^2 k^2$ and $\xi(k,t)$ is the Fourier
transformed white noise with autocorrelation
\begin{equation}
K^{\xi}(k,t,k',t')
=2 \pi \sigma^2
\delta(k+k') \delta(t-t').
\end{equation}
Equation (\ref{GOFourier}) defines an OUP for each $k$.
It has the general solution
\begin{eqnarray}
\label{GOFourierGeneralSolution}
& &\varphi(k,t) = e^{-c(k) t / \tau} \varphi(k,0) \nonumber \\
& &+ \frac{1 }{ \tau} e^{-c(k) t / \tau} \int_0^t ds \xi(k,s) e^{ c(k) s / \tau}.
\end{eqnarray}
Stationarity and homogeneity is ensured if the initial values have the 
autocorrelation
\begin{equation}
K^{\varphi}(k,0;k',0)
=\frac{\sigma^2}{2 \tau}\frac{2 \pi}{c(k)}\delta(k+k'),
\end{equation}
as for the Fourier transform of a spatial OUP with
variance $\sigma^2/4 \tau \lambda$. The autocorrelation function in the
stationary and homogeneous case is
\begin{eqnarray}
\label{GSRcorcontk}
K^{\varphi}(k,t;k',t') & = & \frac{\sigma^2 }{ 2 \tau}
\frac{2 \pi}{ c(k)} \delta(k+k') e^{-c(k) |t - t'| / \tau},
\end{eqnarray}
which is up to  constant factors in accordance with \cite{LB93,VMR94}.
Inverse Fourier transform gives 
\begin{eqnarray}
\label{GOIntegral}
& & K^{\varphi}(x-x',t-t') \nonumber \\
& &= \frac{\sigma^2 }{ 2 \tau}\frac{1}{2 \pi} \int_{-\infty}^{\infty} dk
\frac{1 }{ c(k)} e^{-c(k) |t - t'| / \tau-i k (x-x')},
\end{eqnarray}
where we introduced $K^{\varphi}(x-x',t-t')=K^{\varphi}(x,x';t,t')$. To
calculate the integral on the right hand side of Eq.\ (\ref{GOIntegral}) we
introduce $\tilde{k}=\lambda k$ and $\tilde{c}(\tilde{k})=1+\tilde{k}^2=c(k)$
and note that $K^{\varphi}$ depends only on $\rho=(x-x')/\lambda$ and 
$s=|t-t'|/\tau$. The derivative of Eq.\ (\ref{GOIntegral}) with respect to $s$
reduces to the Fourier transform of a Gaussian
\begin{eqnarray}
& &\frac{\partial K^{\varphi}(\rho ,s )}{\partial s}
= -\frac{\sigma^2 }{ 2 \tau \lambda}\frac{1}{2 \pi} \int_{-\infty}^{\infty}
d\tilde{k}
 e^{-\tilde{c}(\tilde{k}) s -i\tilde{k}\rho}  \nonumber \\
& & =
-\frac{\sigma^2 }{ 4 \tau \lambda}\frac{1}{ \sqrt{\pi s}}
e^{-s-\rho^2/4 s}.
\end{eqnarray}
Integration with respect to $s$ gives
\begin{eqnarray}
& & K^{\varphi}(\rho ,s) = - \frac{\sigma^2 }{ 4 \tau \lambda}
\frac{1}{ \sqrt{\pi}}
\int_{s_0}^{s} ds' \frac{1}{\sqrt{s'}} e^{-s'-\rho^2/4 s'}   \nonumber \\
& & = - \frac{\sigma^2 }{ 4 \tau \lambda}\frac{2}{ \sqrt{\pi}}
\int_{\sqrt{s_0}}^{\sqrt{s}} dy
e^{-y^2-\rho^2/4 y^2}   \\\nonumber
& & =  - \frac{\sigma^2 }{ 4 \tau \lambda}\frac{1}{2}
\Bigg[
e^{\rho} {\rm{erf}}\left(y+\frac{\rho}{2y}\right)
+
e^{-\rho} {\rm{erf}} \left(y-\frac{\rho}{2y}\right) 
\Bigg]_{\sqrt{s_0}}^{\sqrt{s}},
\end{eqnarray}
where   ${\rm{erf}}\,(x)=\frac{2}{\sqrt{\pi}}\int_0^x dt\, e^{-t^2}$ is the
error function. In the limit $s \rightarrow \infty$ the autocorrelation should
vanish, hence 
\begin{eqnarray}
\label{GOCorrReal}
& & K^{\varphi}(\rho , s) =-\lim_{s_0 \to \infty}\frac{\sigma^2 }{8\tau \lambda}
 \\\nonumber& &\times\Bigg\{
e^{\rho}\left[{\rm{erf}}\left(\sqrt{s}+\frac{\rho}{2 \sqrt{s}}\right)
-{\rm{erf}}\left(\sqrt{s_0}+\frac{\rho}{2 \sqrt{s_0}}\right)\right]\\\nonumber 
& &+e^{-\rho} \left[{\rm{erf}}\left(\sqrt{s}-\frac{\rho}{2\sqrt{s}}\right)
-{\rm{erf}}\left(\sqrt{s_0}-\frac{\rho}{2\sqrt{s_0}}\right) \right] \Bigg\}.
\end{eqnarray}
The limit $s_0 \to \infty$ should be carefully taken. If we are interested   in
the limit $\lambda \to 0$ or in the asymptotics for  large $\rho$ the
corresponding operation has to be done before $s_0 \to \infty$.
The limit $\lambda \to 0$ of Eq.\ (\ref{GOCorrReal}) leads to  
\begin{equation}
\label{lambdatozero}
K^{\varphi}(s, x-x')=\frac {\sigma^2}{2\tau}\delta (x-x') e^{-s}  
\end{equation}
as to be expected.  Evaluating first the limit $\tau \to 0$ of Eq.\
(\ref{GOCorrReal}) results in 
\begin{equation}
\label{tautozero}
K^{\varphi}(t-t', \rho)=\frac {\sigma^2} {4\lambda} \delta (t-t')(1+|\rho|)e^{-|\rho|}. 
\end{equation}
Independent on the order of the limits we obtain for
both $\lambda$ and $\tau \to 0$ the result for spatiotemporal Gaussian white
noise $K^{\varphi}=\sigma^2 \delta (x-x')\delta (t-t')$ which can be also 
directly infered from Eq.\ (\ref{GO}).

The asymptotics for $\rho \gg 1$ and $s={\rm const}\neq 0$ and, alternatively,
for $s \gg 1$ and $\rho={\rm const}$, is obtained employing 
${\rm{erf}}(z) \sim \pm 1 - \frac {1}{\sqrt{\pi}z}\;e^{-z^2}$ for $z \to \pm
\infty$, cf.\ e.g. \cite{AS}, as
\begin{eqnarray}
\label{largerho}
K^{\varphi}(\rho , s) \sim
\frac{\sigma^2 }{ 4 \tau \lambda} {\sqrt \frac{s} {\pi}} 
\frac {e^{-\rho^2/4s}}{s-\rho^2/4s}\;e^{-s}.
\end{eqnarray}

For $s \ll 1$ and $\rho={\rm const}\neq 0$ one obtains from Eq.\
(\ref{GOCorrReal}) after first doing $s_0 \to \infty$ and employing again
the asymptotics of ${\rm{erf}}(z)$
\begin{eqnarray}
\label{smalls}
K^{\varphi}(\rho , s) \sim
\frac{\sigma^2 }{ 4 \tau \lambda}\Bigg\{e^{-\rho} + {\sqrt \frac{s} {\pi}} 
\frac {e^{-\rho^2/4s}}{s-\rho^2/4s}\;e^{-s}\Bigg\},
\end{eqnarray}
where the second term on the right hand side vanishes for $s\to 0$.

For $\rho \ll 1$ and $s={\rm const}\neq 0$ expanding ${\rm{erf}}\left(\sqrt{s}
\pm \rho/{2 \sqrt{s}}\right)$
and $e^{\pm \rho}$ one obtains from (\ref{GOCorrReal}), independent on
the order of the limits, up to second order in $\rho$
\begin{eqnarray}
\label{smallrho}
K^{\varphi}(\rho , s) \approx
\frac{\sigma^2 }{ 8 \tau \lambda}
\Big\{\left(1\!-\!{\rm{erf}}\sqrt{s}\right)
\left(2\!+\! \rho^2\right)\! -\! \frac{\rho^2}{\sqrt{\pi s}}e^{-s}
\Big\}.
\end{eqnarray}
The limit $\tau \to 0$ leads to $K^{\varphi}(\rho , s)=\sigma^2/(4 
\lambda)\delta (t-t')\left(1-\rho^2/2 \right)$ in accordance with the
expansion of Eq.\ (\ref{tautozero}) for small $\rho$. Using the asymptotics
$1-{\rm{erf}}\sqrt{s} \sim e^{-s}(1-1/2s)/\sqrt{\pi s}$ for large $s$ one
obtains $K^{\varphi} \sim  \sigma^2 /(4 \tau \lambda \sqrt {\pi s}) e^{-s}
(1-\rho^2/4s)$ which agrees with the expansion of Eq.\ (\ref{largerho}) for
small $\rho$.

The autocorrelation function should solve the equation obtained by
multiplying Eq.\ (\ref{GO}) with $\varphi(x',t')$ and averaging,
\begin{eqnarray}
\label{GOCorrFunctionDGL}
&& \tau \frac{\partial }{ \partial t} K^{\varphi}(x-x',t-t') = -
\left( 1 - \lambda ^2 \Delta \right)K^{\varphi}(x-x',t-t')\nonumber \\ 
&&+\Theta (t'-t)\frac{\sigma^2 }{2 \tau \lambda}\frac{1}{\sqrt{\pi s}}
e^{-s-\rho^2/4s},
\end{eqnarray}
which is fulfilled by (\ref{GOCorrReal}). In the limit $\lambda \to 0$ the
inhomogenity reduces to that of (\ref{inhDGLKOUP}) multiplied by $\delta(x-x')$
as it should be. In the limit $\tau \to 0$ the inhomogeneity of Eq.\
(\ref{GOCorrFunctionDGL}) becomes $\sigma^2/(2\lambda) \delta (t-t')
e^{-|\rho|}$ \cite{help50} which can be also directly derived. 

Sancho et al.\ \cite{SGG98} claimed that the decay of correlations is
exponentially dominated in both space and time. The above results show that
this is generally not the case for $d=1$, see \cite{ExpDecay}.

\subsection{Spatially Discretized Scheme}

Garc\'\i a-Ojalvo et al.\ \cite{GSR92} calculated the autocorrelation of the GSR
process for 
$d=2$ in discrete space. Here we repeat the procedure in $d=1$ to compare it
with the continuous case. 
The spatially discretized version of Eq.\ (\ref{GO}) reads
\begin{eqnarray}
\label{GODiscreteReal}
\tau \frac{\partial }{ \partial t}\varphi_{j}(t) = -\varphi_{j}(t)  +
\lambda^2 \Delta \varphi_{j}(t) + \xi_{j}(t),
\end{eqnarray}
where the Euler discretization of the Laplacian is
\begin{equation}
\Delta \varphi_{j}(t)=\frac{1 }{ \Delta x^2} \left(\varphi_{j+1}(t)-2 \varphi_{j}(t)+
\varphi_{j-1}(t) \right).
\end{equation}
In discrete space we have to rescale the white noise
according to
\begin{equation}
\langle \xi_{j}(t) \xi_{j'}(t')\rangle=\frac{\sigma^2 }{ \Delta x} 
\delta_{j,j'} \delta(t-t').
\end{equation}
Again, as in the continuous case, we Fourier transform, solve the
decoupled equations and calculate the autocorrelation
function. We define the discrete Fourier transform
on the spatial lattice as
\begin{equation}
\varphi_{\mu}(t)=\Delta x \sum_{j=0}^{N-1} e^{i (2 \pi / N ) \mu j} \varphi_{j}(t).
\end{equation}
Hence the inverse Fourier transform is given by 
\begin{equation}
\varphi_{j}(t)=\frac{1}{N \Delta x} \sum_{\mu=0}^{N-1} e^{-i (2 \pi / N ) \mu j} \varphi_{\mu}(t).
\end{equation}
Greek indices are used in Fourier space and latin indices in real space. The
indices run from $0$ to $N-1$ in both real and Fourier space; due to
periodic boundaries $-\mu$ has to be interpreted as $N-\mu$. Now we can Fourier
transform Eq.\ (\ref{GODiscreteReal})
\begin{equation}
\label{GODiscreteFour}
\tau \frac{\partial }{ \partial t} \varphi_{\mu}=
-c_{\mu} \varphi_{\mu}(t)+  \xi_{\mu}(t),
\end{equation}
where
\begin{equation}
c_{\mu}=1-2 \frac{\lambda^2}{\Delta x^2}
\left[ \cos\left(\frac{2 \pi \mu}{N} \right)-1\right].
\end{equation}
The autocorrelation function of the Fourier transformed
white noise is
\begin{equation}
\langle \xi_{\mu}(t) \xi_{\mu'}(t')\rangle=\sigma^2 N \Delta x
\delta_{\mu,-\mu'} \delta(t-t').
\end{equation}
As in continuous space, Eq.\ (\ref{GODiscreteFour}) defines
an Ornstein-Uhlenbeck process with autocorrelation time
$\tau / c_{\mu}$ for each $\mu$. The stationary autocorrelation
can be computed in complete analogy to continous space as the
inverse Fourier transform of
\begin{equation}
\label{GSRcorrrec}
K^{\varphi}_{\mu,\mu'}(t-t')
= \frac{\sigma^2}{ 2 \tau }
\frac{N \Delta x}{c_{\mu}} \delta_{\mu,-\mu'} e^{-c_{\mu} {|t-t'| / \tau}}.
\end{equation}
Hence the stationary autocorrelation in discrete space is
\begin{equation}
\label{GOCorrDiscr}
K_{j-j'}(t-t')=\frac{\sigma^2 }{ 2 \tau} \frac{ 1 }{ N \Delta x} \sum_{\mu=0}^{N-1}
\frac{1}{ c_{\mu}} e^{-c_{\mu} {|t-t'| / \tau}-i (2 \pi /N) \mu (j-j')}.
\end{equation}
Since $c_{\mu}=c_{N-\mu}$ 
the imaginary part of the sum
vanishes. 
For $|t-t'|/\tau \gg 1$ the
autocorrelation is dominated by the first term $e^{-|t-t'|/\tau}$ in the sum
($\mu=0$). However, this is not so for $|t-t'| \approx \tau$, cf.\ Fig.\
\ref{GO1dAnalytNumer}b. 

Observing $N \Delta x = L$, $L$ being the system size, and identifying $k=2
\pi \mu /L $ we have in the  limit $\Delta x \to 0$ the correspondence $c_{\mu}
\to c(k)=1+\lambda^2 k^2$. Hence Eq.\ (\ref{GSRcorrrec}) corresponds to
(\ref{GSRcorcontk}) and Eq.\ (\ref{GOCorrDiscr}) to (\ref{GOIntegral}) after
doing the limit $L \to \infty$ in an  appropriate way.

\subsection{Comparison with Numerics}

The initial conditions for a stationary field in Fourier space were chosen as
independent Gaussian random numbers with variance $\sigma^2 (N \Delta x)^2/\tau
c_{\mu}$ for each $\mu$. The spatial autocorrelation was computed using the
correlation theorem (cf., e.g., \cite{Press93}) valid for weak stationary ergodic
processes
\begin{equation}
{\cal F} \left[ \langle g(x_0) g(x_0+x)\rangle \right] =
{\cal F}[g(x)] {\cal F}[-g(x)],
\end{equation}
where ${\cal F}[g(x)]$ denotes the Fourier transform of $g(x)$.
The procedure is faster in numerical simulations and
gives the same results as the real-space approach, moreover
the inverse Fourier transform can be avoided if one is only
interested in spatial correlations.

\begin{figure}[t]
\centering
\scriptsize
\psfrag{a}{\bf (a)}
\psfrag{b}{\bf (b)}
\psfrag{0.5}{0.5} \psfrag{1.0}{1.0} \psfrag{0.0}{0.0}
\psfrag{0}{0} \psfrag{32}{32} \psfrag{64}{64}
\psfrag{x}{$x-x'$}
\psfrag{t}{$t-t'$}
\psfrag{K}[][][1][180]{$K^{\varphi}(x-x',0)$}
\includegraphics[totalheight=4.5cm]{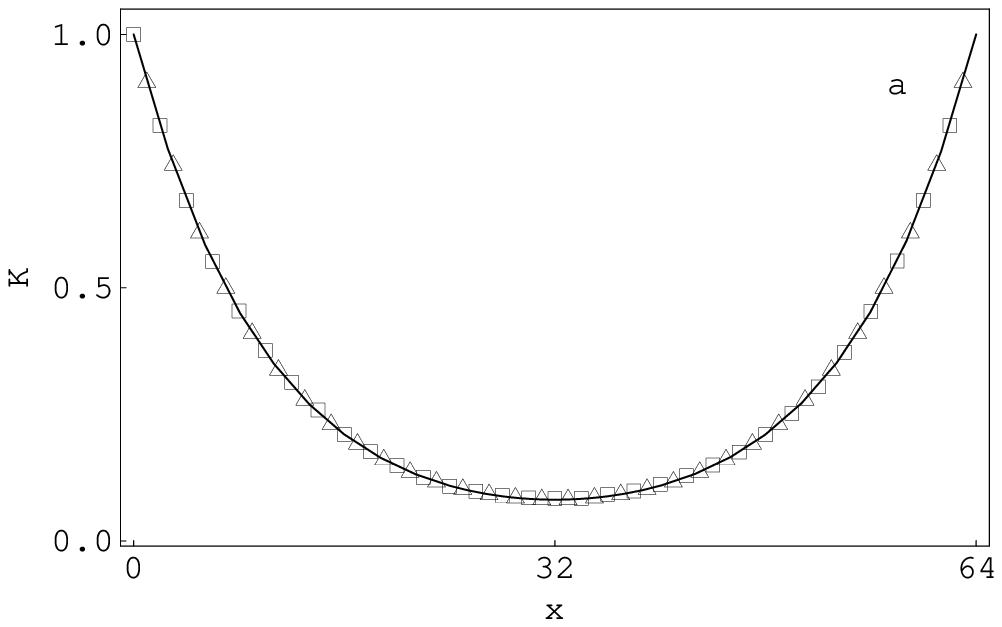}
\hfill
\psfrag{K}[][][1][180]{$K^{\varphi}(0,t-t')$}
\includegraphics[totalheight=4.5cm]{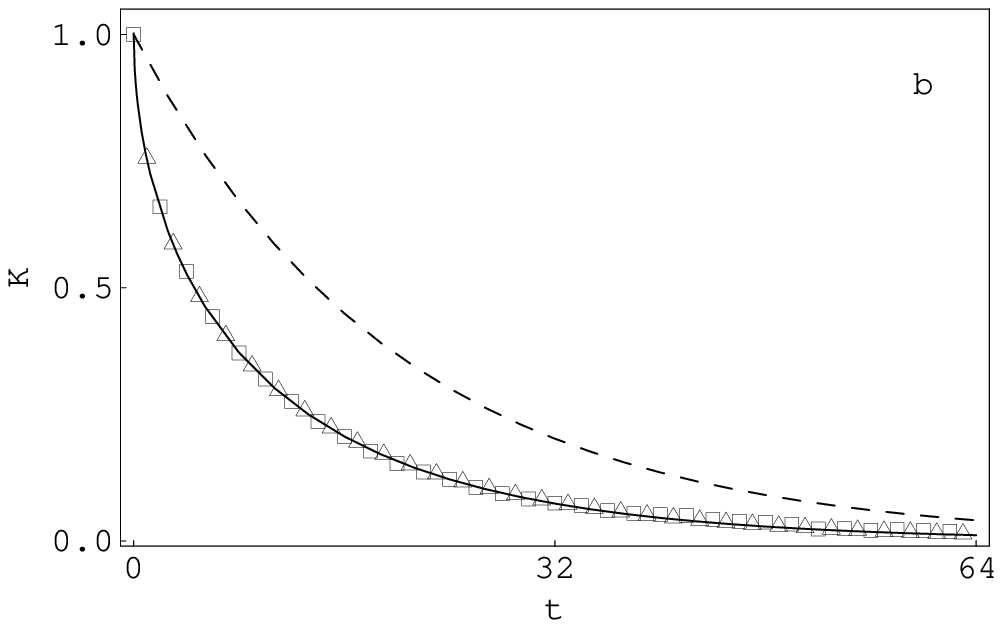}
\caption{Autocorrelation (normalized  by the variance) of the GSR process  in 1+1
dimensions. Comparison of simulations in Fourier space with analytical results.
(a) shows the spatial dependence after a transient period $t=t'=1000$ (note the symmetry
due to periodic boundary conditions), and (b) the temporal dependence for
$x=x'=32$. 
Analytic results from the continuous approach (solid
line), Eq.\ (\ref{GOCorrReal}), and from the discrete approach (triangles), Eq.\
(\ref{GOCorrDiscr}), practically coincide with numerical data
(squares) obtained with Eq.\ (\ref{GO}). The dashed line in (b) shows for
comparison the
autocorrelation of a temporal OUP with $\tau=20$.
Averages over $10^5$ realizations ($N=64$, $\tilde{\sigma}=1$,
$\Delta x = \Delta t = 1$, $\lambda=10$, $\tau=20$).}
\label{GO1dAnalytNumer}
\normalsize
\end{figure}

Figure \ref{GO1dAnalytNumer} compares numerical and analytical
results for the GSR process in 1+1 dimensions.

\begin{figure}[here]
\centering
\scriptsize
\psfrag{0.5}{0.5} \psfrag{1.0}{1.0} \psfrag{0.0}{0.0}
\psfrag{y}[][][1][180]{$K^{\varphi}(0,t-t')$}
\psfrag{t}{$t-t'$} \psfrag{x}{$x-x'$}
\psfrag{0}{0} \psfrag{5}{5} \psfrag{10}{10} \psfrag{15}{15}
\psfrag{20}{20} \psfrag{25}{25}
\includegraphics[totalheight=4.5cm]{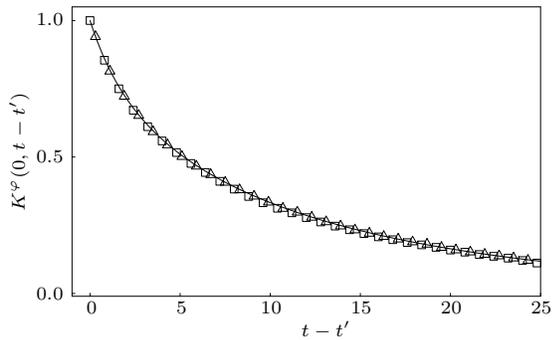}
\caption{Autocorrelation (normalized  by the variance) of the GSR process in 1+1
dimensions. Comparison of simulations in real space for $x=x'=128$ with periodic boundary
conditions (triangles) and  without periodic boundary conditions (squares).
Both practically coincide with the analytical result, Eq.\ (\ref{GOCorrDiscr}). 
Averages  over $10^5$
realizations after a transient period   $t'=500$
($N=256$, $\tilde{\sigma}=1$, $\Delta x =1$,  $\Delta t = 0.1$, 
$\lambda=2$, $\tau=20$).}
\label{GOBCSim}
\normalsize
\end{figure}
Simulations in real space give results which coincide with those in
Fourier space; we refrain from demonstrating this here.
A simulation in real space has the disadvantage that the maximal
possible time step $\Delta t$ is restricted by
$\Delta t < \tau \Delta x^2 / 4 \lambda^2$, otherwise the discrete
Eq.\ (\ref{GODiscreteReal}) looses stability, cf.\ \cite{GSR92}.

Another issue is the dependence on boundary conditions which is shown in Fig.
\ref{GOBCSim}. Since in Fourier space we always have periodic boundary
conditions we work in real space, where the above stability condition enforces
to use a smaller $\Delta t$. In real space simulations without periodic
boundary conditions one has to impose a stochastic boundary process. We used a
temporal OUP with $\tau=20$ which has a different autocorrelation than the
temporal autocorrelation of the GSR process. Therefore we show data after a
transient period apart from the boundaries of the system. Clearly, all
procedures give numerical data matching very well with the analytical result.

\section{Conclusions}
\label{Conclusions}

We introduced in 1+1 dimensions a  spatiotemporal stochastic process with an
autocorrelation exponentially  decaying both in space and time, thus being a
generalization of the OUP.  An analogous  generalization to  higher spatial
dimensions, although formally possible, seems physically not meaningful: The
autocorrelation function should not factorize in the spatial variables. 

The situation resembles to that of the checkerboard process in 1+1 dimensions 
\cite{Feynman,GJKS84,Ibison99,PS01} driven by a velocity changing randomly the
sign which is modeled by the simplest discrete process with exponentially
decaying autocorrelation, the dichotomous Markovian process. The checkerboard
process is intimately connected with the Dirac equation or the
Klein-Gordon equation in $d=1$. Also there, the generalization to higher
spatial dimensions meets nontrivial difficulties \cite{GJKS84,Ibison99}. \\

\acknowledgments 
The work was partially supported by the DFG (grant Be 1417/3). A.T.
acknowledges support by the  Studienstiftung des Deutschen Volkes.
Thanks is due to Dr. Markus Brede for a valuable remark.

\small
\bibliographystyle{plain}

\begin{thebibliography}{99}
\bibitem{help50}        Observe that  $1 /(2 \tau \sqrt{\pi s})
			e^{-s-\rho^2/4s} \to \delta(t-t')e^{-|\rho|}$ in  the
			limit $\tau \to 0$, cf.\ \cite{GH} formula 314.9a.
\bibitem{ExpDecay}	For $t > t'$ an exponential decay of the temporal
			correlation with $e^{-|t-t'|/\tau}$  would imply a
			constant spatial correlation, which is clearly
			unphysical. Similarly, a spatial decay of the
			autocorrelation with $e^{-|x-x'|/\lambda}$ would lead
			to a temporally constant autocorrelation.
\end{thebibliography}

\end{document}